\documentclass[aps,twocolumn,epsf,floats,pre,nofootinbib]{revtex4}
\usepackage{graphics,graphicx,epsfig}
\usepackage{amssymb,color}
\usepackage{epsf,epstopdf,wrapfig}
\usepackage{amsmath}
\usepackage{amsmath}
\usepackage{amssymb}
\usepackage{graphicx}
\usepackage[colorlinks=true, allcolors=blue]{hyperref}
\usepackage{bm}
\usepackage{xr}
\usepackage{scrextend}

\usepackage[export]{adjustbox}

\usepackage{lipsum}

\begin{document}
\title{Marginal speed confinement resolves the conflict between correlation and control \\ 
in collective behaviour}

\author{
Andrea Cavagna$^{1,2,3}$, 
Antonio Culla$^{1,2,*}$,
Xiao Feng$^{1,2}$,
Irene  Giardina$^{1,2,3}$, 
Tomas S. Grigera$^{1,4,5,6}$,
Willow Kion-Crosby$^{1,2}$,
Stefania Melillo$^{1,2}$,
Giulia Pisegna$^{1,2}$,
Lorena Postiglione$^{1,2}$,
Pablo Villegas$^{7,8}$
}
\affiliation{$^1$ Istituto Sistemi Complessi, Consiglio Nazionale delle Ricerche, UOS Sapienza, 00185 Rome, Italy}
\affiliation{$^2$ Dipartimento di Fisica, Universit\`a\ Sapienza, 00185 Rome, Italy}
\affiliation{$^3$ INFN, Unit\`a di Roma 1, 00185 Rome, Italy}
\affiliation{$^4$ Instituto de F\'\i{}sica de L\'\i{}quidos y Sistemas Biol\'ogicos CONICET -  Universidad Nacional de La Plata,  La Plata, Argentina}
\affiliation{$^5$ CCT CONICET La Plata, Consejo Nacional de Investigaciones Cient\'\i{}ficas y T\'ecnicas, Argentina}
\affiliation{$^6$ Departamento de F\'\i{}sica, Facultad de Ciencias Exactas, Universidad Nacional de La Plata, Argentina}
\affiliation{$^7$ IMT Institute for Advanced Studies, Piazza San Ponziano 6, 55100 Lucca, Italy}
\affiliation{$^8$ “Enrico Fermi” Research Center (CREF), Via Panisperna 89A, 00184 - Rome, Italy}
\affiliation{$^*$ Corresponding author, e-mail: a.culla@uniroma1.it}

\begin{abstract}
\section*{Abstract}
\noindent
Speed fluctuations of individual birds in natural flocks are moderate, due to the aerodynamic and biomechanical constraints of flight. Yet the spatial correlations of such fluctuations are scale-free, namely they have a range as wide as the entire group, a property linked to the capacity of the system to collectively respond to external perturbations. Scale-free correlations and moderate fluctuations set conflicting constraints on the mechanism controlling the speed of each agent, as the factors boosting correlation amplify fluctuations, and vice versa. 
Here, using a statistical field theory approach, we suggest that a marginal speed confinement that ignores small deviations from the natural reference value while ferociously suppressing larger speed fluctuations, is able to reconcile scale-free correlations with biologically acceptable group's speed. We validate our theoretical predictions by comparing them with field experimental data on starling flocks with group sizes spanning an unprecedented interval of over two orders of magnitude.
\end{abstract}

\maketitle

\section*{Introduction}

\noindent
Since the early stages of the effort to formulate a mathematical description of collective behaviour, the fundamental dynamical rule common to most theoretical models has been that of  local mutual imitation: each individual within the group tends to adjust its state of motion to that of its neighbours \cite{reynolds1987flocks, heppner1990stochastic, huth1992simulation, vicsek+al_95, couzin+al_02, Chate_2008, romanczuk2012swarming, grossmann2013self}. This type of imitative behaviour can be either explicitly prescribed by the model through a direct interaction between the particles' velocities \cite{reynolds1987flocks, huth1992simulation, vicsek+al_95, couzin+al_02}, or it may be an effective interaction emerging from simpler positional rules, as attraction and repulsion \cite{heppner1990stochastic, romanczuk2012swarming, grossmann2013self, perna2014duality}, depending on the coarse-graining level we decide to work at. In either case, effective imitation of the local neighbours is the cornerstone of self-organised collective dynamics. 
The early models also assumed that all individuals within the group moved with the same constant speed \cite{reynolds1987flocks,  heppner1990stochastic, huth1992simulation, vicsek+al_95, couzin+al_02, Chate_2008}. In that case, mutual imitation requires each particle to only adapt the orientation of its velocity to that of its neighbours. However, in real instances of collective behaviour, be they natural or artificial, the individual speeds fluctuate \cite{cavagna+al_10, bazazi+al_2011, kudrolli2008swarming}, hence mutual imitation requires a particle to adjusts also its speed to that of the neighbours. In contrast to orientation, though, speed control cannot be left just to mutual imitation, as nothing then would prevent particles to move in sync at unreasonably large (or small) speeds. One therefore needs to devise a control mechanism aimed at keeping the individual speed of each particle in the ballpark of some reference value, $v_0$, which is set by the biomechanical constraints of a given species in the case of natural collectives, or  by the technological constraints in the case of artificial swarms. While $v_0$ is a system-specific parameter, typically linked to the hardware of the animal or robot, and therefore hard to change, how the individual speeds are allowed to fluctuate around $v_0$ is what actually constitutes the software, namely the essence of control, which is a fundamental problem both for natural \cite{bialek+al_14, hemelrijk2015scale, Fish_linear, niwa1994self} and for artificial collective behaviour \cite{Drones_linear, Vehicles_linear}; fluctuations of the individual speeds cannot just occur randomly, because when the group is under perturbation the speed change of each individual must influence those of other individuals in order to propagate information across the system. Therefore, collective response is inextricably linked to the speed control mechanism.

A vivid exemplification of this connection comes from one of the most spectacular instances of natural collective behaviour, namely starling flocks. Experimental observations have shown that speed correlations in these systems are scale-free: the changes in speed of one individual are statistically connected to those of all other individuals across the flock  \cite{cavagna+al_10}. More precisely, in statistical physics language, what happens is that the spatial range of speed correlations, i.e. the speed correlation length, grows linearly with the group's size: no matter how large is a flock, individual speed changes are correlated to each other. This is a startling phenomenon that does not have an obvious mathematical explanation, and that very likely is essential to an efficient collective functioning of this natural system. In support of this view come the extensive numerical simulations of \cite{hemelrijk2015scale}, which show that scale-free correlations are necessary to grant cohesion to the group, preventing fragmentation under external perturbations. Collective response and propagation of information, and how these traits impact on the evasion manoeuvres that the group can display in the face of external perturbations, are not only concerns of starling flocks, and not even only of natural systems, but of all instances, biological and artificial, of self-organized collective behaviour. Speed control is therefore a crucial issue in both biology and engineering.

\begin{figure}
\centering
\includegraphics{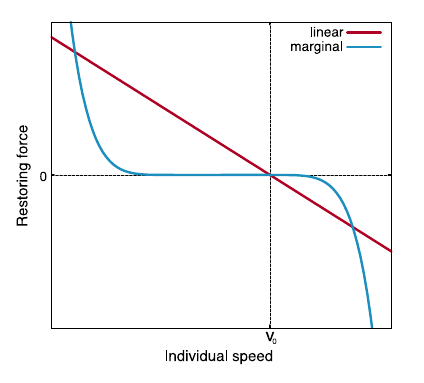}
\caption{{\bf Qualitative sketch of linear vs marginal speed-restoring force.} In the linear case the force pulls the speed back to its natural reference value $v_0$ proportionally to the deviation from $v_0$. Instead, in the marginal case, the force is extremely weak for small deviations from the reference speed, while it increases very sharply for large deviations, harshly suppressing them.
}
\label{fig_1}
\end{figure}

Scale-free correlation is not at all a mild requirement set on speed control and this is demonstrated by the fact that the most widespread speed control mechanism used in the literature, namely linear control, does not work. The simplest way to control speed is indeed through a linear restoring force: whenever the speed $v_i$ of particle $i$ deviates from the natural reference value $v_0$, it gets `pushed back' proportionally to the deviation (Fig.\ref{fig_1}). Linear speed control is widely used to study the collective behaviour of the most diverse systems, from migrating cells \cite{schienbein1993langevin}, bird flocks \cite{bialek+al_14, hemelrijk2015scale}, fish schools \cite{Fish_linear}, and pedestrian collectives \cite{Pedestrian_linear_1, Pedestrian_linear_2, Pedestrian_linear_3}, to robots swarms \cite{Drones_linear}, and vehicle crowds \cite{Vehicles_linear}, to name just a few examples; it lies at the center of most current implementations of collective behaviour. By using field data on starling flocks, numerical simulations of self-propelled particles and statistical field theory, we show here that linear speed confinement entails an intrinsic conflict between yielding a reasonable group's speed and producing scale-free correlations. At its core, the problem is that to reproduce long-range correlations linear control requires a weak speed-confining force, so that the particles' speeds are very loosely confined around their reference natural value, $v_0$; but when this happens, entropic forces push the typical speed of the group to grow significantly larger than $v_0$, not only causing a disagreement between theory and experiments, but more generally giving a severe discrepancy between reference speed and group's speed, which is problematic in any collective system.

We will show that, to resolve this conflict, it is convenient to employ a different speed control mechanism, whose fundamental idea is quite simple: small speed fluctuations elicit nearly zero restoring force, while larger speed fluctuations are pushed back extremely sharply (Fig.\ref{fig_1}). The great advantage of this kind of control is that the low stiffness for small fluctuations boosts correlation, while the sharp increase of the confining force for large fluctuations always grants a plausible speed to the group. For mathematical reasons that will be clearer later on, we call this mechanism, marginal speed control.  Marginal control was first studied on purely speculative grounds in \cite{cavagna2019CRP}, although no self-propelled particles simulations, nor comparison with the experimental data were conducted in that study. Here, we will provide theoretical, numerical and - most importantly - field empirical evidence indicating that a model of collective behaviour based on marginal speed control can produce scale-free correlations and acceptable group's speed without the need to fine-tune any parameters. Although we use starling flocks as a crucial biological benchmark for validation, our results are general, as both numerical results and theoretical calculations support marginal control, irrespective of the specific experimental system one explores. Hence, we argue that marginal speed control may become a centrepiece of collective behaviour at a general level.

\begin{figure}
\centering
\includegraphics{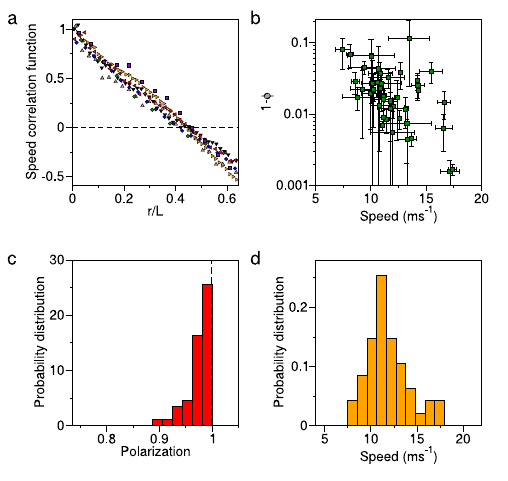}
\caption{{\bf Experimental evidence on starling flocks.} 
{\bf a}: The equal-time space correlation function of the speed fluctuations (for the definition of the correlation function see Methods), plotted against the distance $r$ between the birds rescaled by the flock's size $L$, for some typical flocks (each color corresponds to a different flock); the fact that all the curves collapse onto each other indicates that the spatial range of the speed correlation, namely the correlation length $\xi_\mathrm{sp}$, scales with $L$, i.e. that the system is scale-free (see also Fig.\ref{fig_3}a and Fig.\ref{fig_3}c). 
{\bf b}: Scatter plot displaying polarization vs mean speed of each flock in all recorded events; as the polarization, $\Phi=(1/N) \left|\sum_i \mathbf{v}_i/v_i\right|$, is quite close to $1$ for all flocks, it is more convenient to plot $1-\Phi$ in log scale. Data show that starling flocks are highly ordered systems, incompatible with the standard notion of near-criticality (green points correspond to medians over time, error bars to median absolute deviations). The probability distributions of polarization and mean speed are reported in panels {\bf c} and {\bf d}, showing that the typical mean group's speed is $12$ m$\mathrm{s^{-1}}$, with fluctuations of about $2$ m$\mathrm{s^{-1}}$.
}
\label{fig_2}
\end{figure}

\section*{Results and Discussion}

\vskip 0.5 truecm
\noindent
{\bf Experimental evidence from a natural system}

\noindent
We consider $3D$ experimental data on natural flocks of starlings ({\it Sturnus vulgaris}) in the field.  To the data previously reported by our lab in \cite{cavagna+al_08, cavagna+al_08b, attanasi+al_14}, we added new data from our most recent campaign of acquisition conducted in 2019-2020 (see Methods for details of the experiments, and Table S1 in the SI for all biological data in each acquisition). The new data expand the span of the group sizes between $N=10$ and $N=3000$ animals, a wider interval than any previously reported study. As we shall see, this expansion of the dataset will be crucial in selecting the correct theory.

The three main experimental results that are of interest for us here are the following: 
{\it i)} Speed fluctuations in natural flocks are correlated over long distances, namely their spatial correlation functions are scale-free (Fig.\ref{fig_2}a) \cite{cavagna+al_10}. The connected correlation function measures the similarity between the speed changes of different birds at a certain distance (see Methods for the mathematical definition of connected correlation function); how rapidly this correlation decays with the distance defines the correlation length, $\xi$. In standard systems, normally $\xi$ has a certain fixed value (the scale of correlation), which does not depend on the size of the system, $L$. In scale-free systems, though, one finds that $\xi$ is proportional to $L$, meaning that there is no actual scale of the correlation, apart from that set by the system's size, $L$. Flocks have this very non-trivial property.
{\it ii)} Flocks are highly ordered systems. The polarization, $\Phi=(1/N) \left|\sum_i^N \mathbf{v}_i/v_i\right|$ (where $N$ is the total number of birds in the flock, $\mathbf{v}_i$ is the vector velocity of bird $i$, and $v_i$ is its modulus, i.e. speed), is always quite large, typically above $0.9$ (Fig.\ref{fig_2}b and c). 
The large polarization indicates that these systems are deep into their ordered phase, which rules out the possibility that flocks are close to an ordering transition; hence, near-criticality in the standard ferromagnetic sense cannot be invoked to explain scale-free correlations. Moreover, unlike the case of orientations, scale-free correlations of the speed cannot be explained as the effect of a spontaneously-broken continuous symmetry \cite{goldstone1961field}; in fact, in standard statistical physics, fluctuations of the modulus of the order parameter are heavily suppressed in the ordered phase, so they are very much short-range correlated \cite{patashinskii_book}. 
The origin of scale-free correlations of the speed is therefore far from being trivial. 
{\it iii)} Finally, flock-to-flock speed fluctuations are moderate (Fig.\ref{fig_2}d). The average cruising speed of starlings within a flock is about $12$ meters-per-second (m$\mathrm{s^{-1}}$), with typical fluctuations of $2$ m$\mathrm{s^{-1}}$ \cite{cavagna+al_10}. This is also the typical cruising speed of an entire flock, namely  $s= (1/N) \sum_i v_i$, whose distribution is reported in (Fig.\ref{fig_2}d). Hence, neither the individuals, nor the group, ever cruise at a speed much different from the natural reference value, $v_0$. As we shall see, this seemingly puny experimental trait may become tremendously difficult to reconcile with scale-free correlations at the theoretical and practical level.

\vskip 0.5 truecm
\noindent
{\bf General theory}

\noindent
The reference flocking dynamics we will consider here is one in which the animals' velocities interact through a direct coupling, aimed at describing the effective imitation between neighbouring individuals.
This kind of dynamics can be written in a compact way as follows,
\begin{eqnarray}
\frac{d\mathbf{v}_i}{dt} &=& -\frac{\partial H}{\partial \mathbf{v}_i} + \boldsymbol{\eta}_i 
\label{eq_1} 
\\ 
\frac{d\mathbf{x}_i}{dt} &=& \mathbf{v}_i  \ ,
\label{eq_2}
\end{eqnarray}
where $\boldsymbol{\eta}_i$ is a white noise with strength proportional to $T$, a parameter playing the role of an effective temperature in the statistical physics context, namely $\langle \boldsymbol{\eta}_i(t) \cdot \boldsymbol{\eta}_j(t') \rangle = 2dT\delta_{ij}\delta(t-t')$; $H$ is a cost function (or effective Hamiltonian), whose derivative with respect to $\mathbf{v}_i$ represents the social force acting on the particle's velocity (the effective friction coefficient in front of $\dot{\mathbf{v}}_i$ in \eqref{eq_1} can be set to $1$ through an appropriate rescaling of time \cite{zwanzig_book}). 
In order to implement an imitation dynamics we can use the following - very general - cost function,
\begin{equation}
H = \frac{1}{2}J \sum_{i,j}^N n_{ij} (\mathbf{v}_i- \mathbf{v}_j)^2  + \sum_i^N V(\mathbf{v_i}) \ ,
\label{eq_3}
\end{equation}
where the first term represents the imitation interaction between particles' velocities, having strength $J$, and the second term is the speed control term, which affects each particle independently. The adjacency matrix, $n_{ij}$, is $1$ for interacting neighbours and $0$ otherwise, and the self-propulsion part of the dynamics, \eqref{eq_2}, implies that the interaction network depends on time, $n_{ij}=n_{ij}(t)$. 
The first term in the cost function favours neighbouring individuals to have similar velocity, hence it contains an alignment component of the dynamics that was first explored in the seminal Vicsek model \cite{vicsek+al_95,vicsek_review, ginelli2016physics}; in the Vicsek model, though, the speed of the particles was kept constant, $|\mathbf{v}_i|=v_0$, so that velocity imitation only impacted upon the orientations. Here, on the other hand, we want to study speed fluctuations and their correlations, hence we relax the Vicsek constraint of fixed speed: the first term in $H$ favours mutual imitation of orientation and speed, while the confining potential, $V(\mathbf{v_i})$, keeps the speed of each particle confined around the natural reference value, $v_\mathrm{0}$. Notice also that the Vicsek dynamics was a discrete one, while here we wish to work in the continuum time limit, a generalization far from trivial, which has required a great deal of elaboration since the original Vicsek model was introduced (see, for example, \cite{niwa1994self} and \cite{grossmann2012active}).

 Equation \eqref{eq_3} is certainly not the only possible choice, as more sophisticated cost functions could be considered. 
 Likewise, more complex dynamical evolutions than \eqref{eq_1} \eqref{eq_2} could be adopted, including different kinds of fluctuating terms \cite{grossmann2012active} or correlated noise.  Our aim here is to seek the simplest possible description able to explain the data; we therefore choose  the minimalistic framework of equations \eqref{eq_1}\eqref{eq_2}\eqref{eq_3} as a starting point for the theoretical analysis. Moreover, apart from the simplicity of this modelling approach, it must be said that the analysis of biological data based on statistical inference \cite{bialek+al_12,bialek+al_14,mora2016local} indicates that this class of model provides a good statistical description of natural flocks of birds. We will discuss in the conclusions how models based on separate fluctuations for speed and orientations \cite{romanczuk2011brownian, grossmann2012active} may be relevant for describing other types of biological systems.

The interaction term in the cost function $H$ in \eqref{eq_3} involves the full velocity vectors and therefore regulates mutual adjustment of both speeds and flight directions: individuals who tend to have similar directions also tend to have similar speeds. The actual speed of a bird will then be the product of the interplay between the individual confining force and the collective imitation among the birds. We could have considered two distinct imitation terms for speeds and flight directions, allowing these quantities to be adjusted independently. However, it has been shown in \cite{bialek+al_14} that the two models are substantially equivalent in explaining experimental data on starling flocks, suggesting that a unique adjustment interaction between full velocities already captures most of the relevant information in the data.

\vskip 0.5 truecm
\noindent
{\bf Linear speed control}

\noindent
The simplest control, and indeed the one used in the majority of models with fluctuating speed  to date,  consists of a harmonic potential confining the speed  \cite{schienbein1993langevin, erdmann2000brownian, romanczuk2011brownian, grossmann2012active, bialek+al_14, hemelrijk2015scale}, 
\begin{equation}
V(\mathbf{v}_i)= g \, (v_i-v_0)^2 \ ,
\label{eq_4}
\end{equation}
where $v_i = |\mathbf{v_i}|$. This potential generates a linear restoring force acting on the speed in the equation of motion \eqref{eq_1}, hence it is called {\it linear speed control}. The parameter $g$ is the stiffness of the restoring force, and it can be interpreted as the elastic constant of a spring keeping the speed around its natural reference value, $v_0$. 
The determination of the correlation length $\xi_\mathrm{sp}$ of the speed fluctuations in the case of a linear control has been worked out in \cite{bialek+al_14}, and it gives, 
\begin{equation}
\xi_\mathrm{sp} = r_1  \left(\frac{J n_c}{g}\right)^{1/2} \ ,
\label{eq_5}
\end{equation}
where $r_1$ is the mean inter-particle distance and $n_c$ is average number of interacting nearest-neighbours. The explanation of \eqref{eq_5} is simple:  the theory defined by \eqref{eq_3} and \eqref{eq_4}, has a critical point at $g=0$, where the correlation length diverges. Conversely, large values of the speed stiffness $g$ suppress the range of speed correlations \cite{bialek+al_14}. To have scale-free correlations with linear control one must have have $\xi_\mathrm{sp} \gg L_\mathrm{max}$ (where $L_\mathrm{max}$ is the size of the largest flock in the dataset), which can be achieved by fixing the stiffness $g$ to be much smaller than $1/L_\mathrm{max}^{2}$.

\begin{figure*}[!t]
\centering
\includegraphics{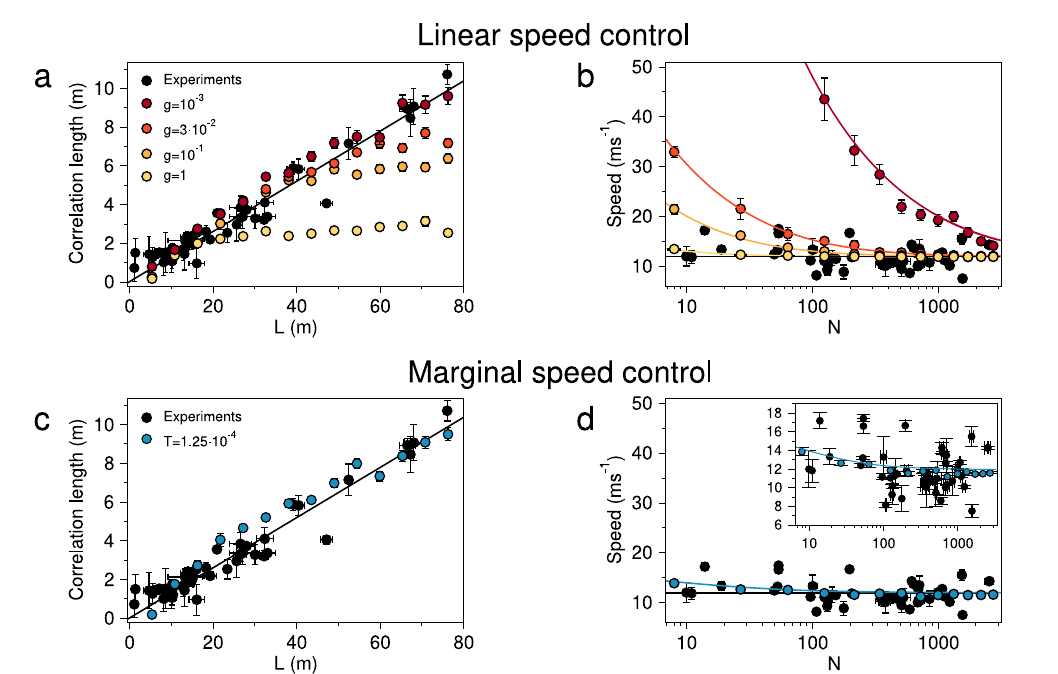}
\caption{{\bf Linear vs Marginal speed control.} 
{\bf a}: Natural flocks show a clear scale-free behaviour of the speed correlation length, $\xi_\mathrm{sp}$, which scales linearly with $L$ (Pearson coefficient $r_P=0.97$, $p<10^{-9}$). SPP simulations with linear speed control yields scale-free correlations over the entire range of $L$ only at the smallest value of the stiffness $g$ (dark red). 
{\bf b}: Natural flocks show no detectable dependence of their mean speed on the number of birds in the flock (Spearman coefficient $r_S=-0.13, p=0.21$; the black line is the average over all flocks). SPP simulations with linear control  give a near-constant speed compatible with experiments only at the largest value of the stiffness $g$ (light yellow); coloured lines represent the theoretical prediction of \eqref{eq_6}. 
Linear speed control is therefore unable to reproduce both experimental traits at the same time. 
{\bf c}:  The correlation length in SPP simulations with marginal speed control scales linearly with $L$ over the full range, provided that the temperature/noise $T$ is low enough to have a polarization equal to the experimental one. 
{\bf d}: At the same value of the parameters as in panel c, SPP simulations with marginal control give mean group's speed very weakly dependent on $N$, fully compatible with the experimental data; the $y$ scale of the main plot is set to make a comparison with panel b, while in the inset we report the same data over a smaller $y$ range to appreciate the agreement between theory (blue line) and simulations, and to show that the trend of group's speed with $N$ of the marginal model is really weak.
[Each length scale $\ell$ pertaining to numerical simulations is reported in meters by normalizing it to the inter-particle distances, $\ell=\ell^\mathrm{sim} (r_1^\mathrm{exp}/r_1^\mathrm{sim})$, where $r_1^\mathrm{exp}$ and $r_1^\mathrm{sim}$ are the mean inter-particle distances of experiments and simulations respectively.
Numerical and experimental correlation lengths are reported on the same scale by matching the curves at the scale-free value of the parameters; numerical and experimental speeds are reported on the same scale by matching the curves at the largest value of $N$. Colored points correspond to averages over numerical data, error bars to standard deviations. Black points correspond to the median (over time) of experimental data for each individual flocking event, error bars to median absolute deviations.]
}
\label{fig_3}
\end{figure*}

This theoretical scenario is confirmed in Fig.\ref{fig_3}a, where we report the correlation length of the speed fluctuations, $\xi_\mathrm{sp}$, vs the system's size, $L$, in numerical simulations of self-propelled particles (SPP) regulated by linear speed control (colored points - see Methods and SI for details of the SPP simulations): when the speed stiffness $g$ is small enough, namely smaller than $1/L_\mathrm{max}^{2}$, the correlation length $\xi_\mathrm{sp}$ scales linearly with $L$ over the whole range (dark red points), thus reproducing the scale-free nature of the experimental correlation length (black points). On the contrary, if $g$ is larger than $1/L_\mathrm{max}^{2}$, the range of the correlation grows linearly with $L$ only up to a certain size, and then it saturates to its bulk value \eqref{eq_5} (orange and yellow points). We conclude that, if correlations were our only experimental concern there would be no need to increase the speed stiffness $g$ beyond $1/L_\mathrm{max}^{2}$, and everything would be fine.  

However, when we turn our attention to the mean speed of the flock, $s = (1/N) \sum_i v_i$, the linear theory becomes problematic. Empirical data show that the mean speed does not change much from flock to flock and it does not have any dependence on the flock's number of birds, $N$ (black points in Fig.\ref{fig_3}b). Let us see what is the prediction of the linear theory for the mean speed, $s$. Calculating the probability distribution, $P(s)$, from equations \eqref{eq_1}-\eqref{eq_2}  is a prohibitive task, due to the time-dependence of the interaction network, $n_{ij}(t)$; however, previous studies have shown that, in the deeply ordered phase in which flocks live, the timescale for rearrangement of $n_{ij}(t)$ is significantly larger than the relaxation time of the velocities, hence one obtains reasonably accurate results by assuming a time-independent form of $n_{ij}$, namely a fixed interaction network \cite{mora2016local}; as we shall see from the perfect agreement between off-equilibrium SPP simulations and theory, this approximation works very well. Under this assumption (plus some more bland algebraic approximations - see SI for details) one can calculate the probability distribution of the mean speed with linear control, obtaining for $d=3$ the result,
\begin{equation}
P(s) = \frac{1}{Z} s^2 \exp\left[- \frac{Ng}{T}(s-v_0)^2\right]  \ ,
\label{eq_6}
\end{equation}
where $Z$ is a normalization factor. We can easily evaluate the peak of this distribution, that is the typical value of the mean speed of the group,
\begin{equation}
s_\mathrm{typical} = \frac{1}{2}v_0 \left(1  + \sqrt{1+\frac{4T}{N g \, v_0^2}} \ \right) \ .
\label{eq_7}
\end{equation}
For $N\to\infty$ we get $s_\mathrm{typical} = v_0$, so all is good for infinitely large groups, as their typical speed is just the same as the natural reference speed, $v_0$. But in finite groups serious troubles emerge, as the typical speed  grows for decreasing $N$, eventually becoming absurdly larger than the natural reference value, $v_0$; and because in \eqref{eq_7} the combination $Ng$ appears, this problematic effect is all the more serious the smaller the speed stiffness $g$, so that for very weak control, even relatively large flocks will have a biologically implausible speed. Yet weak stiffness $g$ is exactly what we need to grant strong correlations! Linear control has therefore a serious problem.

The physical reason for this drift of the mean speed in the linear theory is the following.
In absence of the prefactor $s^2$ in the distribution \eqref{eq_6}, a decrease of the stiffness $g$ would increase the flock-to-flock fluctuations of the mean speed, but its typical value would be always equal to the natural one, namely $v_0$. The $s^2$ prefactor, though, changes this, pushing the maximum of the distribution at larger and larger speed for decreasing $g$. Where is this prefactor from? It is essentially the Jacobian of the change of variable between the $d$-dimensional velocity vector and the modulus of the velocity (see SI); in generic dimension, $s^{d-1} ds$ is the volume in phase space of all configurations with the same mean speed, but variable velocity direction (an identical term appears in the Maxwell-Boltzmann speed distribution). This is an entropic term, which boosts the probability of large speed merely because there are more ways to realise larger rather than smaller velocity vectors. 
When the imitation force is strong (as it is within a flock) and the speed-confining force is weak (as it must be for the sake of scale-free correlation), the system is allowed to gain entropy by increasing in a coordinated fashion all the individual speeds of the particles; as this entropic push is not suppressed by a strong enough exponential decay, it gives rise to unreasonably fast groups.

This theoretical prediction is confirmed by a comparison between numerical SPP simulations ruled by linear speed control and experimental data. Fig.\ref{fig_3}b shows that, once the reference speeds $v_0$ of theory and experiments are matched at the largest sizes, for small values of $N$ and $g$ numerical flocks with linear speed control (dark red points) have a mean speed that is incompatible with that of actual experimental flocks (black points), which shows no appreciable dependence on $N$. To contrast the increase of the  mean speed in smaller SPP flocks one needs a larger value of the speed stiffness $g$ (light yellow points), but this depresses the range of the speed correlations, so that one fails to reproduce scale-free behaviour, Fig.\ref{fig_3}a. This is the blanket-too-short dilemma of linear speed control: either we use a speed stiffness $g$ small enough to reproduce scale-free correlations even at the largest observed values of $N$, but in that case we get implausible large group's speed at low $N$ (dark red points), or we increase $g$ to tame the entropic boost of the speed and keep it within the experimental fluctuations at low $N$, but then we lose scale-free correlations at large $N$ (light yellow points).
Linear speed control cannot yield both experimental traits at the same time. 

The numerical simulations that we present in Fig.3 have been performed in a standard cubic box of side $L$, with periodic boundary conditions; given that the density is $1$ (see Methods), the number of particles in each simulation is $N=L^3$. Natural flocks, on the other hand, do not have a cubic aspect ratio \cite{ballerini+al_08b}, but rather three main axis of different sizes, so that $N=L_1 L_2 L_3$, where the main (i.e. longest) axis $L_1$ is what we define as the flock's linear size, $L=L_1$; flocks are in fact quite elongated, with an aspect ratio $L_1/L_2$ which typically ranges between 6 and 8, depending on the flock. This means that if a flock and a simulation have the same value of $L$, they will not have the same number of particles $N$, and vice-versa. Hence, one possible objection to our discussion about the failure of the linear theory is that the fair comparison should be between experimental elongated flocks and simulations in equally elongated non-cubic boxes, with aspect ratio as similar as possible to the natural one. We run these elongated simulations and we present the results in the SI; what we find is that the results do not change: even in an elongated geometry similar to real flocks, the linear model cannot fit the experimental data. More precisely, if we set the stiffness to $g=1$, in order to have the group's speed under control at all values of $N$ (see Fig.3b), then the correlation is not scale-free, and $\xi$, instead of scaling with $L$, rapidly saturates as in Fig.3a (see SI for the data). The shortcomings of the linear model are therefore not related to the system's geometry, nor to the different ways in which $L$ and $N$ are related to each other.

We have seen that linear speed control is unable to produce scale-free speed correlations and at the same time to keep a limited value of the group's speed, with any fixed value of the stiffness $g$. But what about a tuning mechanism? Could one imagine that $g$ changes with size, in order to keep both experimental traits on board? Let us see this.
In order to have scale-free correlations at all observed sizes, one needs $\xi_\mathrm{sp} \gg L$ for each $L$, a condition that, together with \eqref{eq_5}, leads to,
\begin{equation}
g \ll \frac{a}{L_\mathrm{max}^{2}}
\label{eq_8}
\end{equation}
where $L_\mathrm{max}$ is the size of the largest flock in the dataset and $a=r_1^2 J n_c$ collects all size-independent quantities. On the other hand, from \eqref{eq_7} we see that, in order to have a typical flock's speed reasonably close to the natural reference value, $v_0$, one must ensure that $T/(N g v_0^2)\ll 1$ for all observed sizes; if we use the reasonable approximation $N \sim L^3/r_1^3$, where $r_1$ is the mean nearest neighbour distance, we obtain the condition,
\begin{equation}
g \gg \frac{b}{L^3_\mathrm{min}}
\label{eq_9}
\end{equation}
where $L_\mathrm{min}$ is the size of the smallest flock in the data-set, and where once again we have grouped into the parameter $b=r_1^3 T/v_0^2$ all size-independent constants. Once the spectrum of observed values of $L$ is wide enough, the two bounds \eqref{eq_8} and \eqref{eq_9} cannot be satisfied both with one single value of the speed control stiffness $g$. The only way to reconcile linear speed control with the empirical observations then would be to assume the existence of a tuning mechanism such that the speed stiffness $g$ depends on the size $L$ of the flock according to the condition, 
\begin{equation}
\frac{b}{L^3} \ll g(L) \ll \frac{a}{L^2}
\label{eq_10}
\end{equation}
This is a rather narrow strip for $g(L)$ to live in, so that a biological mechanism fulfilling \eqref{eq_10} would require some very tricky size-dependent fine-tuning. But in fact, even that could be insufficient: the two inequalities in \eqref{eq_10} are asymptotic, namely they require the stiffness $g$ to stay well clear of both boundaries, $1/L^3$ and $1/L^2$, not just between them; for medium-small values of $L$ this becomes harder and harder to achieve. We can think of condition \eqref{eq_10} as a wedge on the $(g,L)$ plane, a wedge that closes rapidly when $L$ decreases; the only way to keep this wedge open also for small values of $L$ would be to tune also all other parameters, $r_1, J, n_c, v_0$ etc, beside tuning the stiffness $g$. Such a fine tuning seems unlikely, if not impossible, to achieve. We believe it is more realistic to turn to some other, tuning-free, control mechanism.

\vskip 0.5 truecm
\noindent
{\bf Marginal speed control}

\noindent
In statistical physics, the correlation length $\xi$ is connected to the inverse of the quadratic curvature of the (renormalized) potential, calculated at its minimum \cite{le1991quantum, goldenfeld_lectures_1992,binney_book}; very small curvature implies very large correlation length, so that a divergent $\xi$ is always due to a zero second derivative (or marginal mode) along some direction of the (renormalized) potential. This is also the case for linear speed control \eqref{eq_4}: the second derivative of the quadratic potential along the speed is proportional to $g$, hence when $g$ is small, the correlation length is large. The problem, however, is that because the function is quadratic, by decreasing $g$ we weaken the whole speed-confining potential, not just its curvature, hence giving a freeway to the entropic boost we have discussed before, ultimately resulting in the implausible large speed of small flocks.

This state of affairs suggests that we must turn to a confining potential that  does not vanish entirely when its curvature does. To find this potential we proceed through general considerations of symmetry and common sense. First, the potential must keep the speed around the reference natural value $v_0$ and it must diverge for large values of the speed; secondly, it must be rotationally symmetric in the whole velocity vector; third, it must have the simplest mathematical form compatible with the previous conditions and with the experimental evidence. The most general form of a rotationally symmetric potential that confines the speed around the natural reference value $v_0$ is, $V(\mathbf{v_i}) =  \left( \mathbf{v}_i\cdot\mathbf{v}_i-v_0^2\right)^p$, where the integer power $p\geq 2$ must be even, in order to produce a minimum of the potential at $v_0$. 
For $p=2$ we have the classic $\mathrm{O}(n)$ quartic potential of standard vector ferromagnets \cite{patashinskii_book,le1991quantum}, namely,
$V(\mathbf{v_i}) =  \left( \mathbf{v}_i\cdot\mathbf{v}_i-v_0^2\right)^2$; several works that study collective behaviour with variable speed indeed
use this kind of quartic potential \cite{niwa1994self, erdmann2000brownian, d2006self, hanke2013understanding}. Although this may seem a speed control mechanism genuinely different from the harmonic one that we considered in the previous section (after all, it generates a cubic, rather than linear, speed-confining force), in fact it is not. As we have seen, in highly polarized and coherent flocks individual speed fluctuations are relatively mild, so that for $v_i\sim v_0$ we can rewrite the quartic potential as, $V(\mathbf{v_i}) =  ( v_i+v_0)^2 (v_i-v_0)^2 \sim 4v_0^2 (v_i-v_0)^2$, which is nothing else than the harmonic potential that we already took into consideration. Hence, the $p=2$, or quartic, theory is not suitable for our purposes, because it has a non-zero quadratic expansion around $v_0$, which gives a non-zero curvature, leading to a saturation of the correlation length, and therefore to a violation of the scale-free phenomenology. The only way to avoid this would be to put a vanishing amplitude (i.e. stiffness) $g$ in front of the whole quartic potential; but this is exactly what we tried doing in the case of linear control, and it does not work, because then the mean speed of the group explodes. Incidentally, the fact that the `harmonic' or quadratic speed control theory is essentially identical to the quartic $O(n)$ model, shows that that theory is not `harmonic' at all in the actual vectorial degrees of freedom, $\mathbf{v}_i$.

The next simplest possibility is $p=4$, which gives the following speed-control potential \cite{cavagna2019CRP},
\begin{equation}
V(\mathbf{v}_i) =  \frac{1}{v_0^6} \lambda \left( \mathbf{v}_i\cdot\mathbf{v}_i-v_0^2\right)^4
\label{eq_11}
\end{equation}
where, thanks to the $v_0^{-6}$ normalization, the amplitude $\lambda$ has the same physical dimensions as the other coupling constants, $J$ and $g$. The crucial feature of the potential in \eqref{eq_11} is that its second derivative with respect to the speed is always zero, irrespective of the value of the amplitude $\lambda$, hence we will call this marginal speed control. Higher order powers in the expansion of the potential are nonzero and very steep, though, thus confining the speed much more effectively to its reference value, $v_0$, compared to linear control.  Correspondingly, the speed-restoring force is very weak for small deviations from $v_0$, but very strong for large deviations, as one can qualitatively see from the marginal speed-restoring force in Fig.1. We will discuss later about the biological plausibility of this kind of nonlinear speed confinement; here this is for us merely the simplest confining potential, beyond the harmonic one. Although from a physical point of view simplicity is certainly appealing, it is not always the best path to interpreting the experimental results, especially in biological systems, whose complexity often defies the physicist's desire of (over?) simplification; hence, we cannot exclude that other mechanisms, possibly within more complex theories, may equally successfully explain the empirical evidence. 
For example, in the spirit of \cite{huth1992simulation} different kinds of noise could be adopted in the equations, affecting the speed differently than the orientations (like the active noise considered in \cite{grossmann2012active}) or with non-trivial temporal correlations. Moreover, the reference speed $v_0$ might itself be a dynamical variable: even though we do not observe temporal trends of the mean group speed in the data, at least on the scale of our experiment, it could be a relevant issue for other instances of collective motion \cite{bazazi+al_2011}.
Once again, our criterion here is to explore the simplest mathematical possibility that may work.

\begin{figure*}[!t]
\centering
\includegraphics{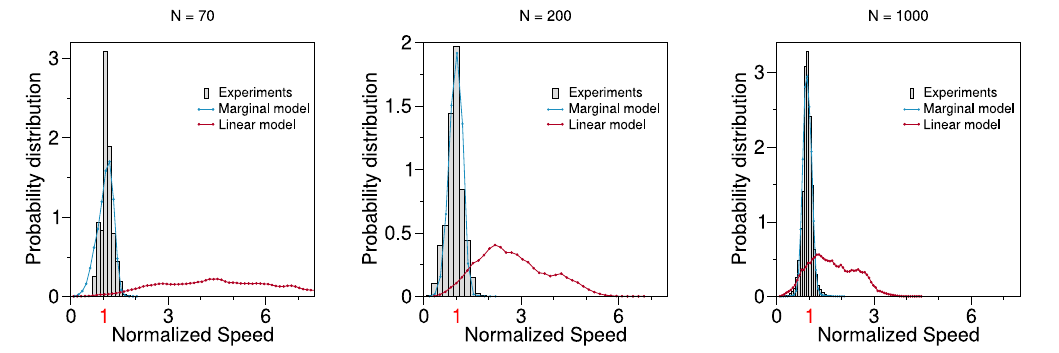}
\caption{{\bf Single particle speed distributions.} 
We measured the single particle speed distribution (as opposed to the mean speed of the group of Fig.\ref{fig_3}), for different groups sizes, for real flocks (grey), the linear model (dark red) and the marginal model (light blue). The parameters in the marginal case are the same as in Fig.3, while for the linear case we fixed the stiffness at $g=10^{-3}$, which is the only value giving scale-free correlations at all sizes (see Fig.\ref{fig_3}). The individual speed on the abscissa has been normalised to the reference speed value $v_0$, in order to compare all cases on the same plot. The data show that linear speed control is unable to fit the experimental distribution even at the largest $N$, and it is a total disaster in the medium and smallest $N$ cases. On the other hand, the marginal theory fits the data in a rather satisfactory way at all values of $N$.
}
\label{fig_4}
\end{figure*}

The complete absence of a quadratic term in the expansion of \eqref{eq_11} seems to suggest that the marginal potential gives rise to an infinite correlation length of speed fluctuations under all physical conditions; in fact, this is not the case. Speed correlations are regulated by the confining potential (i.e. the energy), but also by the fluctuations induced by the noise (i.e. the entropy): at very low noise the marginal potential dominates, so that speed fluctuations are indeed scale-free, while by increasing the noise the correlation is increasingly suppressed by entropic fluctuations \cite{cavagna2019CRP}. In field theory terms, what happens is that at finite temperature entropy provides a non-zero second derivative of the renormalized potential, i.e. a non-zero mass of speed fluctuations, and therefore a finite correlation length. The renormalized curvature only goes to zero at $T=0$, where the speed correlation length diverges.  As a consequence, the marginal theory has a zero-temperature (or zero-noise) critical point, where the correlation length of the speed fluctuations diverges.
A mean-field analysis \cite{cavagna2019CRP} shows that the speed correlation length diverges as,
\begin{equation}
\xi_\mathrm{sp} \sim \frac{1}{T^{1/2}} \ ,
\end{equation}
where the generalized temperature $T$ is the strength of the noise in \eqref{eq_1}.
This scenario has an interesting and very convenient consequence: in the marginal theory, by simply decreasing the noise strength $T$, we bring home two out of three empirical traits, that is a large polarization and a large correlation length, a somewhat unusual result within standard statistical systems, which normally are the less correlated the more they are ordered (we are talking about connected correlation, of course).  But what about the crucial constraint of producing a biologically reasonable group speed?

The calculation of the distribution of the mean speed of flocks under the marginal potential is more complicated than in the linear case (see SI), but under some reasonable approximations one obtains,
\begin{align}
P(s)=\frac{1}{Z} s^{2}\exp{\left[-\frac{N\lambda}{Tv_0^6}(v_0^2-s^2)^4\right]}
\end{align}
which has two great differences compared to the linear case: first, the power four in the exponential tames the entropic push of the $s^2$ term extremely sharply; secondly, the amplitude $\lambda$, unlike $g$, does not need to be small to grant scale-free correlations, so the exponential weight remains always effective in suppressing large values of the mean speed, $s$.
The maximum of this distribution, i.e. the typical mean speed of the flock, is given by (see SI for details),
\begin{align}
s_\mathrm{typical} \simeq 
\begin{cases}
v_{0}\ \ \ \ \ &\textrm{for} \ N \gg \frac{T}{\lambda v_0^2}\\
v_{0} \left(\frac{T}{4N\lambda v_0^2}\right)^{1/8}\ \ \ \ \ \ \ \ \ \ \ \ \ \ \ &\textrm{for} \ N \ll \frac{T}{\lambda v_0^2}
\end{cases}
\end{align}
As in the linear case, for $N\to\infty$, the typical speed of the flocks becomes the same as the natural reference speed, $v_0$, while it increases for smaller sizes. However, in the marginal case this growth is very moderate indeed: the strength $T$ of the noise is small (to have large correlation length) and the amplitude $\lambda$ is finite, so that the crossover size $N=T/(\lambda v_0^2)$, below which the mean speed increases, is very small, thus shielding this regime; moreover, the exponent of the speed's increase for small $N$ is $1/8$, significantly smaller than the exponent $1/2$ of linear speed control (see equation \eqref{eq_7}). For these two reasons marginal speed control is so much more effective than linear control at small $N$. 

These theoretical results are fully confirmed by numerical simulations of SPP flocks regulated by marginal speed control  (Fig.\ref{fig_3}c and Fig.\ref{fig_3}d). Moreover, the comparison with experimental data on real starling flocks in the field indicates that marginal control gives satisfactory results: with just one reasonable set of parameters - the only crucial one in fact being the low noise strength, $T$ - numerical simulation of SPP flocks with marginal speed control reproduce the experimental data very well: the correlation length $\xi_\mathrm{sp}$ scales linearly with $L$  up to the largest size (Fig.\ref{fig_3}c), and the mean speed $s$ shows only an extremely weak increase at low $N$, well within the scatter of the empirical data (Fig.\ref{fig_3}d). 

There is one further - and final - empirical test that we can make to compare the two theories, linear vs marginal; it consists in investigating the individual speeds, and how they fluctuate in the experimental data, compared to the two models. This is a complementary view to the analysis of the mean speed of the group that we presented in Fig.\ref{fig_3}. In Fig.\ref{fig_4} we report the single particle speed distributions for natural flocks, and compare it with the numerical simulations of the linear and marginal model; we do this for three different values of the number of particles $N$. The stiffness in the linear case has been fixed to a value granting scale-free correlations at all sizes (see Fig.\ref{fig_3}a), while the parameters of the marginal model are the same as in Fig.\ref{fig_3}. It is quite clear that individual speed fluctuations are far too large in the case of linear control, even in the largest $N$ case, and they are completely out of scale in the medium-small $N$ cases. On the contrary, marginal control gives very reasonable distributions at all sizes, in very good agreement with experiments.

We stress once again that all three pieces of phenomenology -- large polarization, large correlation length, moderate speed at all group sizes -- are achieved with marginal control by doing just one very sensible thing, namely pushing the system into the ordered phase real flocks naturally belong to. The entropy-triggered conflict between scale-free correlation and moderate group speed that hinders linear control is therefore resolved by the marginal theory without any fuss.

\vskip 0.5 truecm
\noindent
{\bf Biological significance}

\noindent
The highly non-linear marginal potential implies that small speed fluctuations elicit nearly zero restoring force, while larger speed fluctuations are pushed back extremely sharply, in contrast with the constant slope of a linear confining force, Fig.\ref{fig_1}. In bird flocks, small speed fluctuations are not prevented by biomechanical constraints, but they could be depressed by energetic expenditure concerns, as changing the speed requires extra energy consumption; however, starlings prove to be very liberal about their energy expenditure habits while flocking \cite{hamilton1967starling, heppner1974avian, bajec2009organized}: although their metabolic rate is dramatically higher in flight than on the roost \cite{hamilton1967starling}, these birds will spectacularly wheel every day for half an hour before landing, expending energy at a ferocious rate; this suggests that small extra energy expenditures due to small speed fluctuations may indeed be weaker-than-linearly suppressed. On the other hand, large speed fluctuations clash against biomechanical and aerodynamic constraints, which are set very stringently by anatomy, physiology and physics \cite{pennycuick1986mechanical, rayner1988form, rayner1996biomechanical}; therefore, a stronger-than-linear suppression of large speed fluctuations also seems quite reasonable.  
Moreover, we notice that while the linear restoring force is completely symmetric around $v_0$, hence suppressing fluctuations smaller than $v_0$ as strongly as those larger than $v_0$, the marginal force is asymmetric, suppressing decelerating speed fluctuations less harshly than accelerating ones (see Fig.1). It seems reasonable to expect that for birds, as well as for animals in general, accelerating should be harder than slowing down; hence, the asymmetry in the speed-restoring force adds to the biologically plausibility of marginal control.

Even though we tested our conclusions on the case of starling flocks, it may be that marginal control is one simple way (possibly not the only one) to reconcile data with theory in other biological systems, although of course this should be tested experimentally. Field experiments reporting the dynamical trajectories within animal groups are quite rare, especially in three dimensions, and beyond starlings the only data are for pigeons ({\it Columba livia}) \cite{freeman2011group, nagy2013context, watts2016misinformed}, jackdaws ({\it Corvus monedula}) \cite{jolles2013heterogeneous, ling2019behavioural, ling2019collective} and chimney swifts ({\it Chaetura pelagica}) \cite{evangelista2017three}. Monitoring the mean speed of these groups should be straightforward, while checking the speed correlations may be somewhat more laborious, as only a study of correlation at different group's size $L$ would reveal whether or not scale-free correlations are present also in these systems; and yet, this seems to us an essential step to establish on a firmer basis the connection between speed control and correlation, which is the cornerstone of our results.

One interesting issue raised by the experimental studies on jackdaws \cite{jolles2013heterogeneous} and pigeons \cite{freeman2011group, nagy2013context, watts2016misinformed} is the fact that real biological flocks are heterogeneous: birds are old and young, male and females, and beyond these obvious traits one may have more influential individuals in the group, as well as outliers with a non-average behaviour. To what extent the results that we obtained within a perfectly homogeneous model, in which all agents are exactly the same, are robust against heterogeneities? In particular, an important problem is whether or not the deviations from the mean behaviour of a few keystone individuals are acted upon by the rest of the group. Despite the simplicity of the marginal model, we can introduce meaningful heterogeneities in two ways, namely by attributing a different reference speed or a different speed variability to a few individuals. The results (that we discuss in details in the SI) are quite interesting: both the deviation in the reference speed and the deviation in speed variability are indeed tamed by the rest of the group provided that the imitative interaction is strong enough, which is by no means a problem for the marginal model, as stronger interaction means stronger correlation. Hence, not only scale-free correlations and moderate group speed, but also robustness against heterogeneities, are all achieved in the marginal model by simply staying in the strongly-interacting ordered phase. Of course, there are other types of heterogeneities that cannot be studied within our simple model: considering variations in structural size and body mass between individuals within the same species, as well as understanding what happens in mixed-species flocks, require more sophisticated models, as for example the one studied in \cite{hemelrijk2015scale}; and - of course - real experimental studies. The fact that simple heterogeneities are tamed very easily in the marginal model, makes us optimistic about the robustness of marginal control in more general cases.

A further crucial issue to consider when we think about real biological systems, is that the dynamical phases of natural collective behaviour are diverse: starlings' aerial display studied in our data (sometimes called {\it murmurations}) is characterized by a very compact drop-like structure, moving coherently over the roost, often subject to predation \cite{procaccini2011propagating}; chimney swifts display a remarkable circling geometry \cite{evangelista2017three}, while the jackdaws data collected in \cite{ling2019behavioural} display two group-level phases, namely a cruising-to-roost dynamics and an anti-predator mobbing dynamics. It would be helpful to understand what are the properties that these different phases have in common. This is particularly relevant for correlation, which - as we have seen - is a tricky trait to sustain. Consider, for example, correlations in the velocity orientations of animals: statistical physics tells us \cite{goldstone1961field} that when the rotational symmetry is spontaneously broken by the group, namely when out of many equivalent directions of motion only one is selected, scale-free correlations of the orientations emerge automatically in the system. If spontaneous symmetry-breaking seems certainly to be the relevant case for starling flocks swirling over the roost, jackdaws flocks traveling to the roost \cite{jolles2013heterogeneous, ling2019behavioural, ling2019collective} and homing pigeons \cite{freeman2011group, nagy2013context, watts2016misinformed} need to follow one specific direction, hence there is no spontaneous symmetry breaking; similarly, in the case of migrating birds, when longer duration flight carry the individuals along one well-defined route, there is no spontaneous symmetry breaking. In these cases, correlations of the orientations could be quite different from those of starlings. However, speed requires a different correlation mechanism than orientation, as no physical reason automatically grants long-range correlation; hence, it would be really helpful to investigate the link between correlation and speed control in different phases. Our impression is that long-range speed correlations are essential to propagate information in all phases of collective motion, in order to keep a good degree of cohesion in the face of natural variations of the individual speeds; we therefore expect that the interplay between speed control and speed correlation is a very general concern of collective motion. But only experiments can confirm this.

Experimental data on two-dimensional collective motion are somewhat more accessible than $3D$ data. The recent study about sheep herds \cite{ginelli2015intermittent} presents a case where exactly the same interplay between correlation and speed control could be at work, and the presence of intermittency - with its huge speed fluctuations - could make even more urgent the issue of speed control; again, obtaining correlations at different groups size could require some nontrivial work, but the two-dimensional nature of the systems makes the tracking somewhat simpler than in the case of bird flocks. Intermittent motion and large speed distributions have also been observed in locust swarms \cite{bazazi+al_2011}. Similarly, $2D$ data of fish schools in shallow water have been studied for a long time, both numerically \cite{huth1992simulation} and experimentally \cite{HerbertRead:2011p12915, macgregor2020information}; as in the case of herds - and unlike the case of flocks - speed fluctuations in fish schools are substantial, hence the issue of speed control can be quite different than in bird flocks. In fact, there may exist a rather profound difference between animals that cannot change much their speed, as birds within a flock, for which not only very large speeds are forbidden, but also very low ones, because of the very aerodynamics of flight, and animals that can reduce considerably their speed, down to halting, as mammal herds or fish schools, at least to a certain extent; apart from a significantly larger asymmetry in the control mechanisms, the very possibility to reduce the speed to zero could be a game changer. As we have explained above, the entropic push that blows the group speed in the linear case is due to the Jacobian factor $s^{d-1} ds$, which is not kept under control by the exponential factor in the speed distribution; this factor implies that the probability to have speed equal to zero is always vanishing, a property that is certainly true in the case of bird flocks (see Fig.4), but it is not so for organisms that can actually halt. To study such a case, hence, a different kind of model than the one studied here would be required. One possibility would be to control separately the noise fluctuations of velocity orientation and speed, along the lines of the studies in \cite{romanczuk2011brownian, grossmann2012active}, in such a way to eliminate the Jacobian factor and shift the distribution towards lower speed. This would be an interesting direction to explore, provided - of course - that some new measures are taken to grant the second key ingredient we have been investigating here, namely long range speed correlations.

Finally, beyond the horizon of broadening our study to include biological systems other than starling flocks, experiments on artificial self-organized swarms would be the next essential step forward. Artificial swarms would allow first to assess at the embodied level (which is quite different from the numerical one) to what extent group's cohesion depends on the range of the correlation: the fact that long-range correlations (quite rare a condition in physical systems) are so frequent in biological collective behaviour - from bird flocks \cite{cavagna+al_10}, to midge swarms \cite{attanasi2014finite} and down to bacterial clusters \cite{chen2012scale} - has prompted biophysicists to connect this trait to collective response and cohesion, with both theoretical \cite{mora+al_11} and numerical backup \cite{hemelrijk2015scale}; yet, biology is one thing, engineering another one, and no matter how much inspiration the latter takes form the former, it is crucial to check whether the link between correlation and response holds at the technological level. Secondly, in artificial collectives it would be possible - at least to some extent - to tune the reference agents' speed $v_0$ and to tweak the manner individual speed can fluctuate around $v_0$ in a controlled way, which is impossible to achieve in natural systems. Having the possibility to operate on both arms of the problem (correlation and speed control) would be invaluable, given the growing technological relevance of self-organized collective behaviour.

\section*{Methods}

\vskip 0.5 truecm
\noindent
{\bf Experiments.} Empirical observations of starling flocks  have been performed in Rome, from the terrace of Palazzo Massimo alle Terme, in front of a large roosting site at the Termini Railway Station. The experimental technique used is stereoscopic photography, where multiple synchronized video-sequences of a flocking event are acquired from different observation points with a calibrated multi-camera video-acquisition system \cite{Cavagna2015error}. Digital images are then analysed with a specifically designed tracking software \cite{attanasi2015greta}, in order to extract from the raw data the three-dimensional trajectories of the individual birds in the flock.  

Data have been collected across the years during several experimental campaigns. The very first data were collected in the context of the {\it Starflag} project \cite{cavagna+al_08,cavagna+al_08b}, between 2007 and 2010, using Canon D1-Mark II cameras, shooting interlaced at 10 frames-per-second (FPS), with a resolution of $8.2$ Megapixels (MP). A second campaign took place between 2011 and 2012, using faster cameras, namely IDT M5, shooting at $170$FPS with a resolution of $5$MP. A final campaign took place in the last months of 2019 and in January and February 2020. This campaign uses state-of-the-art IDT OS10-4K cameras, shooting at $155$FPS with a resolution of $9.2$MP. Overall, we have data from flocks with sizes ranging between $10$ and $2500$ birds, a span that is essential to differentiate between linear and marginal speed control.

All campaigns have been conducted with a three camera system exploiting trifocal geometry \cite{hartley2004multiple}. The image analysis - segmentation of individual birds, stereometric matching and dynamical tracking - have been performed using the method of \cite{cavagna+al_08,cavagna+al_08b} for the first campaign,  and the most advanced method of \cite{attanasi2015greta} for the second and third campaigns. We summarise in Table S1 in the SI all the experimental quantities used in our analysis for each flocking event.

\vskip 0.5 truecm
\noindent
{\bf Correlation functions and correlation length.}
The speed spatial connected correlation function is defined as \cite{cavagna2018physics}
\begin{equation}
	C(r) = \frac{\sum\limits_{i,j}^N \delta v_i  \ \delta v_j \ \delta ( r - r_{ij})}{\sum\limits_{i,j}^N \delta (r-r_{ij})} \label{eq_15}
\end{equation}
where $N$ is the number of individuals in the system,  $r_{ij}=| \boldsymbol{r}_i -\boldsymbol{r}_j|$  is the mutual distance between individuals $i$ and $j$, and 
\begin{equation}
	\delta v_i = v_i - \frac{1}{N} \sum\limits_k^N  {v}_k
\end{equation}
is the fluctuation of the individual speed $v_i=|\boldsymbol{v}_i|$ with respect to the mean speed of the group $s=(1/N)\sum_k v_k$, evaluated at a given instant of time, thus neglecting effects of common external environmental factors. The function $C(r)$ represents the instantaneous average of mutual correlations among all pairs at distance $r$: in systems with local, distance-dependent interactions, for large enough system sizes, this quantity is a good proxy of the typical correlation at that distance, as computed with the correct theoretical measure. A full discussion of definition \eqref{eq_15}, its asymptotic limit, finite size effects, and behaviour in known cases, can be found in  \cite{cavagna2018physics}. Here we notice that the correlation function \eqref{eq_15} is the only possible definition applicable to experimental data, where no a priori information is available on the true nature of the dynamics. This definition has indeed been used in all the previous analysis of speed correlations mentioned in this paper.  
We display in Fig.S5 of the SI the correlation function \eqref{eq_15} computed, respectively, from experimental data (panel a), and from numerical simulations with a linear speed control model (panel b) and a marginal speed control model (panel c).

For each configuration of the system (at a given time) we estimate the correlation length as:
\begin{equation}
	\xi = \frac{ \int\limits_0^{r_0} \ \textrm{d} r \ r \ C(r)}{ \int\limits_0^{r_0} \ \textrm{d} r \ C(r)} \label{eq_17}
\end{equation}
and then we perform a time average of this quantity over different configurations. The point $r_0$ is the first point at which the correlation vanishes, $C(r_0)=0$;  such a point always exists due to the very definition of correlation function given in \eqref{eq_15} \cite{cavagna2018physics}. Definition \eqref{eq_17} provides a reliable estimate of the correlation length in every regime, both when the system is scale-free with long-range correlations and when the system is far from criticality with short-range correlations (in the linear speed control model we can see all this phenomenology by changing the parameter $g$, see Fig. \ref{fig_3}a). The reason is that  \eqref{eq_17} makes use of the information encoded in the zero-crossing point $r_0$, together with the shape of the entire function $C(r)$.

Different definitions of the correlation length are possible, but one has to be careful, as it is very difficult to find alternative definitions that provide a reasonable estimate of the spatial range of the correlation both in the scale-free case and in the short range case. For example, in the original work on scale-free correlations in flocks \cite{cavagna+al_10}, we used the zero of the correlation function, $r_0$, as an estimate of the spatial span of the correlations, because it scales properly when the system is scale-free, correctly identifying the size of the correlated domains, and scaling linearly with the system's size, $L$; however, $r_0$ is not a good estimate of the correlation length when a system is not scale-free, namely when there is an intrinsic short-range length-scale, $\xi$, as in that case one finds $r_0 \sim \xi\log L$ \cite{cavagna2018physics}. Conversely, in the case of short-range correlations it is relatively easy to determine the correlation length, for example via an exponential fit, while this procedure is unfeasible in the scale-free case, when the correlation function does not have a short-range decay (see Fig. S5 in the SI).

On the other hand, definition \eqref{eq_17} works in all regimes. In the case of a short-range correlation, for example exponential, ($C(r) \sim e^{-r/\hat{\xi}}$), it is easy to verify that \eqref{eq_17} gives $\xi \sim \hat{\xi}$. Conversely, when correlations are scale-free, regardless of the precise shape of the $C(r)$ one has $r_0 \sim L$, and equation \eqref{eq_17} gives a correlation length that scales with $L$ as well; more precisely, for scale-free correlations as the ones we find in flocks, namely with an almost linear decay, \eqref{eq_17} gives, $\xi = r_0/3$; considering that in \cite{cavagna+al_10} we found $r_0 = L/3$, one has that in scale-free flocks, $\xi = L/9$, which is consistent with the data we report in Fig.3. We believe that this factor $3$ correction with respect to $r_0$ is inessential, as the only relevant information in the scale-free case is not the actual value of the span of the spatial correlation, but rather the fact that it scales with $L$.

\vskip 0.5 truecm
\noindent
{\bf Numerical simulations of SPP flocks.}
To investigate the flocking dynamics described by \eqref{eq_1} and \eqref{eq_2}, we perform numerical simulations with a system of self-propelled particles. The flock is modeled as a set of particles moving in a three-dimensional space with update rules for positions and velocities, which are a discretized version of \eqref{eq_1} and \eqref{eq_2}. Following a simple Euler integration scheme \cite{rapaport2004}, we get
\begin{align}
	&\boldsymbol{v}_i(t+\Delta t)= \boldsymbol{v}_i(t) + \Delta t \, \boldsymbol{F}_{i} + \delta\boldsymbol{\eta}_i
	\label{eq_18} \\
	&\boldsymbol{x}_i(t+\Delta t)= \boldsymbol{x}_i(t)+ \Delta t \, \boldsymbol{v}_i(t) 
\end{align}
Here the force $\boldsymbol{F}_i = \boldsymbol{F}_{int} + \boldsymbol{F}_{sc}$ acting on particle $i$ contains both an alignment term $\boldsymbol{F}_{int}$,
\begin{equation}
	\boldsymbol{F}_{int} = - J\sum\limits_j^{N} n_{ij}(t) \left( \boldsymbol{v}_i(t) -\boldsymbol{v}_j(t) \right)
\end{equation} 
and a speed control term $\boldsymbol{F}_{sc}$, which can be either linear:
\begin{equation}
	\boldsymbol{F}_{sc} =2 g \frac{\boldsymbol{v}_i}{|\boldsymbol{v}_i|} \left( v_0 - |\boldsymbol{v}_i| \right)
\end{equation}
or marginal,
\begin{equation}
	\boldsymbol{F}_{sc} = \frac{8 \lambda}{v_0^6} \boldsymbol{v}_i (v_0^2-\boldsymbol{v}_i^2)^3
\end{equation}
The last term in \eqref{eq_18} is a white Gaussian noise with zero mean and variance:
\begin{equation}
	\sigma_{\eta}^2 = 2dT  \Delta t
\end{equation}
where $d=3$ is the space dimension and $T$ is the effective temperature. The matrix $n_{ij}(t)$ is the adjacency matrix that defines which pairs interact;
its entries can assume only the values $0$ and $1$, according to a rule of interaction that can be  metric (i.e. $n_{ij} \neq 0$ if and only if $r_{ij} < r_c$) or topological (i.e. $n_{ij} \neq 0$ if $j$ is one of $i$'s first $n_c$ neighbours) \cite{ballerini+al_08, bialek+al_12}. When working at fixed average density and in the very low temperature region where density fluctuations are small, there is not great difference between metric and topological interaction. Even though natural flocks are known to have topological interactions \cite{ballerini+al_08, bialek+al_12}, we therefore decide to perform simulations with the metric rule, which are much less expensive computationally. In this way, we are able to study systems in $d=3$ with  $N$ up to  $3 \times 10^5$ particles. We consider a metric connectivity matrix with interaction radius $r_c=1.2$, such that the number of nearest neighbours at the time $t=0$ is $n_c=6$, close to the biological value \cite{ballerini+al_08,bialek+al_12}.  We then check a posteriori that the system remains spatially homogeneous in time by computing the distribution of the number of nearest neighbours for every simulation, and verifying that it is always sharply peaked around the initial value $n_c=6$. All the simulations are made in a cubic box (of linear size $L$) with periodic boundary conditions. Individuals are initialized in a global polarized configuration on a cubic lattice with lattice spacing (i.e. nearest neighbour distance) $r_c=1$ and then evolve off-lattice according to rules \eqref{eq_18}. The effective temperature clearly drives the system from a disordered to an ordered state through a phase transition at fixed density. However, since we are considering self-propelled particles, the same configurations can be reached using another control parameter defined as the ratio between the mean first neighbour distance $r_1$, which directly depends on the density of the system, and the interaction radius $r_c$ but at fixed noise. We decide to perform all the simulations at constant density $\rho = 1$, maintaining $r_1/r_c$ constant and choosing the temperature according to the desired polarization.

We  choose the value of the reference speed of the particles $v_0$ and of the integration step $\Delta t$ to ensure an average displacement $\Delta r \simeq v_0 \Delta t$ much smaller than the size of the box $L$. In this way there is a weak rewiring of the interaction network during the time of simulation, consistently with the quasi-equilibrium condition of natural flocks \cite{mora2016local}. The integration step is selected as the maximum value granting a robust numerical integration in terms of errors and stationarity of the system's energy (absence of trends in time or in size). In simulations with marginal speed control this algorithmic stability is achieved with  $\Delta t = 0.01$, while  linear speed control requires a $\Delta t = 0.001$.
Every simulation consists in a run of length $N_{steps}=2 \times 10^4$ steps for thermalization and in an independent run long  $N_{steps}=1.2 \times 10^6$ steps. From the latter  we extract configurations every $1000$ steps in order to compute the quantities needed by our analysis. In Table S2 of the SI we report the values of the other parameters used in the simulations.

\section*{Data availability}
\noindent
The flocks data that were analysed in this study are provided in the Supplementary Information file (Table S1).

\section*{Acknowledgements}
This work was supported by ERC Advanced Grant RG.BIO (785932) to ACa, and ERANET-CRIB grant to ACa and TSG.  TSG was also supported by grants from CONICET, ANPCyT and UNLP (Argentina). The authors acknowledge several illuminating discussions with William Bialek regarding speed control in flocks. ACu wishes to thank Victor Martin-Major for careful advice on the numerics. ACa warmly thanks Frank Heppner for reading the original manuscript and for a decade-long conversation on collective avian behaviour.

\section*{Author contributions}
ACa, IG and TSG designed the study. XF, WKK, SM, LP, and PV - coordinated by SM - collected the experimental data. SM carried out the 3D dynamical tracking of the raw experimental data. ACa, ACu, and TSG performed the analytic calculation. ACu and GP carried out the numerical simulations. ACu analysed the experimental and numerical data. ACa and IG wrote the manuscript.

\section*{Competing interests}
The authors declare no competing interests.

\end{document}